\documentclass[12pt,preprint]{aastex}
\newcommand{\phunit}{photons cm$^{-2}$ sr$^{-1}$ s$^{-1}$ \AA$^{-1}$ }
\newcommand{\egunit}{ergs cm$^{-2}$ s$^{-1}$ \AA$^{-1}$}
\newcommand{\apollo}{{\em Apollo~17} }
\newcommand{\voyager}{{\em Voyager} }

\slugcomment{Astrophysical Journal, {\em in press}, 2002 January 3} 
\shorttitle{The Local Interstellar Radiation Field}
\shortauthors{Richard Conn Henry}
\begin{document}
\title{The Local Interstellar Ultraviolet Radiation Field}

\author{Richard Conn Henry\altaffilmark{1,2} }
\affil{Center for Astrophysical Sciences,\\
Henry A. Rowland Department of Physics and Astronomy,
\\ The Johns Hopkins University,
    Baltimore, MD 21218}

\email{henry@jhu.edu}

\altaffiltext{1}{Director, Maryland Space Grant Consortium.}
\altaffiltext{2}{Member, Principal Professional Staff, The Johns Hopkins University
Applied Physics Laboratory.}

\begin{abstract}
I have used the Hipparcos Input Catalog, together with Kurucz model stellar
atmospheres, and information on the strength of the interstellar 
extinction, to create a model
of the expected intensity and spectral distribution of the 
local interstellar ultraviolet radiation field, under various 
assumptions concerning the albedo $a$ of
the interstellar grains.  (This ultraviolet
 radiation field is of particular interest because of the fact 
 that ultraviolet radiation is capable of profoundly 
 affecting the chemistry of the interstellar medium.)  By comparing my models with
 the observations, I am able to conclude 
 that the albedo $a$ of the interstellar grains
 in the far ultraviolet is very low, perhaps $a=0.1$.  I also advance arguments
 that my present determination of this albedo is {\em much more reliable} than any of
 the many previous (and conflicting) ultraviolet interstellar grain albedo determinations.
 Beyond this,  I show that the 
 ultraviolet background radiation
 that is observed at high galactic latitudes {\em must be extragalactic} 
 in origin, as it cannot be backscatter
 of the interstellar radiation field.
\end{abstract}


\section{Introduction}
The physics of the local interstellar medium is profoundly influenced by the 
intensity and spectrum of the local interstellar radiation field, particularly in the ionizing ultraviolet; 
but our knowledge of the {\em strength} of that radiation field 
has considerable uncertainty.  There are few observations, and existing models are far from perfect.

In the present paper I improve on the method for estimating the brightness of the radiation field
 that was invented by Henry (1977), who showed that by simply integrating
the flux that is expected from the stars that are included in the Yale 
Bright Star Catalog, one could provide a useful estimate
of the local interstellar ultraviolet radiation field intensity.  Henry 
showed that, in the ultraviolet, such an integration {\em converged},
demonstrating that the Bright Star Catalog went sufficiently deep to provide a complete result.

Henry's approach has the serious defect that, as he implemented it, it does 
not predict the {\em scattered-light} component of the interstellar
ultraviolet radiation field.  It is that defect that I repair in the present paper.

There is another reason for wanting to know the brightness and spectrum of the interstellar radiation field
in the ultraviolet, quite apart from its influence on interstellar chemistry.  Henry (1991, 1999) has claimed that the spectrum of 
diffuse ultraviolet background radiation that is observed at high galactic latitudes has a sharp break near
1216 \AA, with only an upper limit measurable below that wavelength, and a roughly constant spectrum at 300 to 400 \phunit
longward of that wavelength, to 2800 \AA\ (as appears in the spectrum of
 Anderson et al. 1979a,b), and continuing through
the visible portion of the spectrum (Bernstein 1998; Bernstein, Freedman, and Madore 2002,
ApJ, in press), and he has suggested that this 
radiation originates in recombination radiation from the 
baryonic component of the intergalactic medium.

A much simpler and more conservative interpretation is that the light in the ultraviolet at high galactic latitudes
is simply galactic starlight that is reflected by high latitude interstellar dust:  some people have argued that 
there {\em is} plenty of such dust (perhaps $A_V=0.1$, according to the assessment
of Henry and Murthy (1994)) at high
galactic latitudes (as revealed, for example, by IRAS) and if that is so, then the observed decline at the shortest
 wavelengths {\em might} be due to nothing more than
 a decline at short wavelengths in the intensity of the general interstellar radiation field.  (If that view, that there
 is plenty of dust, is
 wrong, and there is in fact little dust at high galactic latitudes, then of course the alternative explanation collapses immediately.)

Henry's argument against this simpler picture has been his citation of the observation of the 
Coalsack nebula by Murthy, Henry, and Holberg (1994),
who found that the bright ultraviolet spectrum of reflecting interstellar dust toward the Coalsack, which they observed
with {\em Voyager}, most definitely does {\em not} decline below 
Lyman $\alpha$.  The difficulty with this argument is that the Coalsack is being illuminated primarily by three stars ($\alpha$~Cru, 
B0.5~IV, $V=1.58$; $\beta$~Cru, B0.5~III, $V=1.25$; and $\beta$~Cen, B1~III, $V=0.61$) that are among the brightest ultraviolet
stars in the sky, and, specifically, that these particular stars are exceptionally hot and are also very close physically to the dust 
that they are illuminating, so that their light can be expected to dominate the scattered radiation.
  Thus, one could argue that, in  this case, the illumination is atypical, and that the general interstellar 
radiation field (which is what actually illuminates
the high-galactic-latitude dust) has a cooler spectrum that does indeed decline sharply below Lyman $\alpha$.   Only by 
determining the actual expected spectrum of the interstellar radiation field 
can one rule out this possibility, and a desire to assess this (yea or nay)
formed a major motivation for the present work.

\section{The Star Catalog Integration}

For the new star catalog integrations, I use the Hipparcos Input Catalog, Version~2 (http://cdsweb.u-strasbg.fr/cgi-bin/Cat?I/196), 
which contains information on 118,209 stars.  For each
of these stars, I extracted the spectral class (as well as information on the luminosity class), the $V$ magnitude, and the 
observed $B-V$ color.  From the spectral class and the estimated luminosity class, I ascertained the {\em intrinsic} $B-V$ color
 $(B-V)_{0}$, and hence obtained $A_V=3.1~E_{B-V}\equiv 3.1 ~[(B-V)-(B-V)_{0}$].  Stars 
 with negative $E_{B-V}$ were assigned $A_V=0$.  Specifically, if a star
 was identified as luminosity I or II, I obtained $(B-V)_{0}$ from Allen (2000, page 388, ``Supergiants"); otherwise, from the same table
 in Allen, but the (B-V)$_{0}$ that is listed under ``Main Sequence."  (Giants; that is, luminosity class III stars;
  are not listed separately in Allen, for the hot stars
 that are of greatest interest.)
 
 I obtained the temperature of each star by ascertaining its effective 
 temperature (as a function of that star's spectral class) according to the prescription of
  Allen (2000, page 151, Table 7.6), for main sequence stars, and (same source), Table 7.8
 for supergiants.  Again, giants have no separate listing, for the hot stars that are of greatest interest; giants were therefore
 treated as main sequence stars for the present purposes.  For the very hottest stars I adopted the following 
 temperatures, adapted from Lang (1992, page 137):  O4, 48,000~K; O5, 44,500~K; O5.5, 43,000~K; O6, 41,000~K; O7, 39,000~K; O7.5, 38,000~K;
 O8, 35,900~K; O9.5, 34,600~K.  The few stars that are identified simply as O were assigned 
 effective temperatures of 50,000~K, the hottest Kurucz model.  At the request of the referee, I re-ran my program (almost 13 hours
 on a Macintosh G4 desktop) to see if it makes any difference if I assign spectral class O9 to these same few stars instead.  The difference
 is detectable in the plot of the resulting integrated spectra, below 1000 \AA; but only barely detectable; these stars are few and faint.
   
 I then used the interstellar extinction (see Appendix A) of Cardelli, Clayton, \& Mathis (1989, CCM)
 to find the extinction $A_\lambda$ at each wavelength $\lambda$ of interest.    The two values 
 of the CCM parameter $R_V = A_V/E_{B-V}$ that I
  used in the present work are $R_V=3.1$ (a value that is appropriate for the general 
  interstellar medium) and $R_V=5$ (appropriate for dense clouds).  This 
  parameter $R_V$ characterizes the {\em shape} of the extinction curve
  as a function of wavelength.
  
  Exactly how bright are the stars?  Theory and observation {\em do} agree, although this agreement was not achieved
   without a long struggle.  That stellar atmosphere
  calculations do accurately predict the observed brightnesses of the stars in the far ultraviolet has now been firmly
  established by Holberg et al. (1982, 1991) and Kruk et al. (1997).  
  
  This agreement made my task simple:  for each star, I could simply find the most closely matching stellar 
  flux model of Kurucz (1992).  I used the Kurucz models fp00k2.pck, which
  may be found at
  
   http://cfaku5.harvard.edu/grids/gridP00/. 
   
   I then accumulated, for all 118,209 stars, the quantities
   
  $F_0/2.5119^{V-A_V}/2.5119^{(1.0-a)A_\lambda}$,
  
   for eleven values of the assumed interstellar grain albedo $a$ (assumed to be 
  independent of wavelength), namely $a$ = 0.0, to $a$ = 1.0 inclusive (in steps of 0.1).  
  
  In my expression, $F_0 = 1030 ~K_\lambda /K_{5510}$ photons cm$^{-2}$ s$^{-1}$ \AA$^{-1}$  is the flux 
  expected from an unreddened $V=0$ magnitude star, the Kurucz model is $K_\lambda$, $V$ is the star's
  observed $V$ magnitude, and $A_V$ and $A_\lambda$ are as described above.  The first part of the
  expression, $V_0/2.5119^{V-A_V}$, entirely eliminates all effects of interstellar extinction, thus
  giving the brightness that the star would have, were there no interstellar medium.  The final factor,
  $/2.5119^{(1.0-a)A_\lambda}$ re-introduces the interstellar medium, at the wavelength in
  question and for a particular value of the assumed albedo, {\em a}.  If $a=1.0$, this factor of course has no
  effect at all:   the extinction is entirely {\em scattering}, and the extincted radiation is simply
  returned to the beam (we assume {\em forward} scattering, but in the end, I will shortly argue; that
   assumption does not affect the result).  On the 
  other hand, if $a=0.0$ the {\em entire} extinction is due to absorption, and that amount of light is 
  permanently removed from the beam.  
  
  Note that the adopted procedure means that multiple scattering is taken
  fully into account, and also that cloud structure in the interstellar medium can have no effect on our result.
  
\section{Results}
   
My first experiment was to confirm the result of Henry (1977) concerning the 
convergence of star catalog integrations, for the
special case $a=0$.
In Figure~1 the result is displayed, for the wavelength range 0 to 5000 \AA.  Henry's 
claim is abundantly confirmed.  (Note also that the {\em apparent} convergence of the star catalog
integration at {\em long} wavelengths is caused only by the incompleteness of the catalog.)

The next step in the investigation was to perform the same calculation for the full range of 
possible albedo values, from zero (as has already been presented in Figure~1) 
to unity.  The complete result is shown in Figure~2 for the case $R_V = 3.1$ (and in Figure~3 for the 
case $R_V = 5$), and is {\em very dramatic}:  if the
albedo of the interstellar grains were actually unity, the interstellar ultraviolet radiation 
field would be enormously higher than it is according
to all measurements.  Our first, and uncontroversial, conclusion is that the albedo of the 
interstellar grains in the far ultraviolet
is less than unity.  

The reader will see from Figure 2 that, consistent with my discussion of my method in the 
last section, there is no question that the stellar ultraviolet starlight scattered from dust is generated in the 
present model; for high values of the albedo, a very large amount of scattered light
is present.  The highest curves would be even higher if the Hipparcos catalog went deeper, so
if the result of the present investigation had been that the grain albedo is high (upper
curves) there would be an additional uncertainty in deducing the correct value for
the albedo because of the incompleteness of the catalog; but, as we will see, the deduced
albedo is low, so the catalog is more than adequate for the task (as has been demonstrated in Figure 1).

Note that the results that are shown in Figures 2 and 3 are entirely independent of the value of the Henyey-Greenstein scattering parameter $g$.
(That parameter characterizes the scattering pattern of reflective grains, with $g=0$ representing
 {\em isotropic} scattering; large positive values (that is, values toward unity)
representing strong forward scattering; and negative values indicating backscattering.)  If the grains are 
strongly forward scattering, my calculations
give the actual radiation that is expected from that direction, but if, for example, $g=0$, then the same amount (statistically) of radiation must be present,
simply coming from another direction. 
  Note that this follows directly from spherical symmetry, and so is only true if we confine our attention to
the radiation field integrated over the entire sky, as I do in the present paper.  Detailed 
models of the predicted distribution of ultraviolet scattered starlight have been presented by Murthy and 
Henry (1995), but those models are not relevant to the present special case.

There is a final technical point that needs to be addressed, having to do not with the
model, but with the observations.  The reader will shortly see that very few observations are referenced in this paper,
and that is because I am concerned with the total radiation field, including dust-scattered starlight.  Most 
ultraviolet observations of the sky have been made with instruments, such as TD-1, that had a very
narrow field of view (slow optics).  Of course such experiments still respond to the diffuse scattered starlight,
but the signal from the diffuse readiation is comparable to, and impossible to separate from,
internal instrumental dark current.  Thus, only instruments with a very large field of view
(fast optics) are capable of measuring the integrated radiation field over
the sky.


Next, I asked the narrow technical question, ``is it necessary to interpolate between Kurucz models to get sufficient accuracy
for my purposes?" I was able to prove
that such interpolation is not necessary.  When I compared the models of Figure~2 with the result of
repeating the calculation that led to what is displayed
in Figure~2, for the two extreme values of the albedo, but this time having the 
program select the {\em hotter} of the two models that are closest in 
temperature to the deduced temperature (from the spectral type) of the star, 
instead of picking the model that is {\em closest} in temperature,
the result was differences no greater than 6\% for the case albedo $a=1.0$, and 2\% for albedo $a=0.0$.  The actual
error resulting from failure to interpolate will be only half that, and is 
smaller, surely, than the uncertainty in estimating the temperature
from the spectral class.  In the light of this result, I eschew interpolation.

By far the largest uncertainty in the whole procedure results from our lack of knowledge of the value of $R_V$ for each star, and even that,
we shall see, is not a serious problem.

\section{Comparison with Observation}

The next step, Figure~4, is to expand our view of part of Figure~2 and compare with the observations.  Note that a more convenient
wavelength scale has been used in Figure~4.  The figure shows models for
assumed values of the albedo of $a=0.0$, 0.4, and 0.8. 

Very few observations exist of the intensity or spectrum of the far ultraviolet interstellar radiation field.  Two rocket observations
(Henry, Swandic, et al. 1977 and Opal and Weller 1984) are discussed in Appendix B.  However, I recommend rejection 
of both of these rocket results, and acceptance instead of the \apollo measurement of
Henry, Anderson, and Fastie (1980), which is shown in Figure~4 as filled black circles.  My preference for the \apollo result is 
based on calibration uncertainty.  Henry, Swandic, et al. explicitly state that they 
expect their absolute calibration to be in error by ``about a factor of 2 either way," and the Opal and Weller photometer was
calibrated using similar techniques in the same laboratory.  In contrast, the \apollo experiment was 
very carefully calibrated by William G. Fastie and Donald F. Kerr (1975), and is claimed to be accurate to $\pm$10\%.  That calibration 
was tested in flight by carrying out spectroscopy of bright stars, the result being
(Henry, Weinstein, Feldman, Fastie, and Moos 1975)  good agreement with the observations of others,
and with model atmosphere calculations.  

The \apollo measurement of the interstellar
ultraviolet radiation field was reported by Henry, Anderson, and Fastie (1980), and appears in Figure~4 as the black circles with
$\pm$25\% error bars, which is my estimate of the {\em systematic} error in their measurement (as judged from Figures 3 and 4 of  
their paper).

There is another important reason for rejecting the higher albedo that can be derived (Appendix B) 
from the Henry et al. rocket flight, and that is that it
is inconsisent with the remarkable and important information that is provided in that paper concerning the distribution of the ultraviolet
starlight on the sky.  In Figure~3 of their paper we see a map of the actual distribution of ultraviolet radiation over the sky.  The
radiation is seen to come predominantly from the Gould belt, and {\em not} from the galactic plane.  
We see that there are regions in the galactic 
plane that are relatively {\em very faint} in the ultraviolet.  This could only occur if the albedo of the interstellar grains is low.

By comparing the \apollo\ observations in Figure~4 with the radiation field that would be expected 
were $R_V = 5$ (Figure~3) we see that the latter would predict
far more radiation than is actually observed.  I conclude that our use of $R_V = 3.1$, which choice
 I have already mentioned is the largest source of uncertainty in
our model, is both appropriate and accurate.

Comparison of the \apollo\ observation and the present models (Figure~4) leads to the firm conclusion that the albedo of the interstellar grains
is less than 0.4, and perhaps is about $a=0.1$.  I also conclude that the albedo {\em rises} longward of about 1500 \AA, as shown by the rise in 
the \apollo\ spectrum (see Figure 4).  Henry, Anderson, and Fastie (1980) concluded that this rise was due to very large numbers of fainter stars contributing
to the interstellar radiation field, but the present results, using the deep Hipparcos catalog, demonstrate that they were wrong in that 
conclusion.  I can conceive of no other
reason for the observed rise in intensity than an actual rise in the interstellar grain albedo.

That this is so, is demonstrated decisively by including in Figure~4 my integration, using the work of Landsman (1984), of the 
brightness of the TD-1 stars as measured by Gondhalekhar, Phillips, and Wilson (1980) (red circle at 6.4 m$\mu$).  Their passband is indicated
by a horizontal ``error bar."  This integration involves only the direct light of the stars, and contains no general diffuse glow.  The
fact that this TD-1 data point falls almost on my lowest integration shows that their measurements include {\em all} stars that contribute
significantly to the interstellar radiation field, and shows once again 
that the \apollo rise in intensity at about the same wavelength {\em must be} due to
diffuse galactic light kicking in at these wavelengths.

Opal and Weller (1984) made the interesting remark that about half their observed flux could be accounted for as the flux from only two dozen
of the brightest stars.  I have tested this idea in Figure~5, which shows the contribution at 965 \AA\ that is 
expected from the 100 brightest (at
975 \AA) of the 118,209 stars 
in the Hipparcos catalog.  And indeed, I find that these 100 stars contribute 87.5\% of the flux.  (I worked at 965 \AA, rather than at 
the effective wavelength 975~\AA\ of the Opal and Weller photometer, because the Lyman $\gamma$ line sharply reduces the flux at 975~\AA).
In contrast, the 100 brightest stars at 1535 \AA\ (Figure~6) only contribute 62.1\% of the total flux, at that wavelength,
 of the catalog integration.

That comparatively nearby bright stars contribute so heavily to the total radiation field contributes support to my assumption of spherical
symmetry in arguing that the results of the present work are independent of 
the true value of the Henyey-Greenstein scattering parameter {\em g}.

The final exercise is to allow my deduced interstellar ultraviolet radiation field to 
impinge on a high-galactic-latitude dust layer of, let us say, $A_V=0.1$,
and see what the shape is of the scattered spectrum---in particular, is there a sharp decline shortward of Lyman $\alpha$?  The result
is shown in Figure~7, for three values of the assumed albedo: $a=0.1$, 0.2, and 0.3.  The figure also includes
 a cartoon of the observed diffuse radiation spectrum
at high galactic latitudes.  The result is absolutely decisive:  no decline at all is predicted below Lyman $\alpha$, under any set of assumptions.

\section{Discussion}
   
Determining the values of the interstellar grain albedo $a$, and of the Henyey-Greenstein scattering parameter $g$ for the interstellar grains,
 has always been a very difficult and controversial undertaking.  The 
present model cannot be used to say anything at all about the value of the scattering parameter $g$, to which the present procedure
 is totally insensitive.
  However, the method does provide a decisive result regarding the albedo,
indicating unequivocally that the albedo is very low, let us say $a=0.1$, in agreement with the determination of Murthy, Henry, and Holberg (1991).  

The enormous scatter that exists among the large number of published values for the 
parameters $a$ and $g$, as reviewed, for example, by Bowyer (1991),
has one fundamental cause, and that is, that, almost without exception, methods for 
evaluating these parameters require an understanding of the
three-dimensional geometry of the system that is being observed and investigated, 
and this typically requires
assumptions that may well  not be justified.  The method that is used in the present paper is unique,
in that the three-dimensional geometry is known with complete precision:  we are in the center, and all the stars that provide the 
radiation are distributed
 radially outward from us---no exceptions.  Furthermore, the present method takes multiple scattering of starlight by interstellar grains 
into account completely.  Finally, our method
is completely insensitive to the spectrum of cloud structure in the interstellar medium.  Thus, the reader should have far greater 
confidence in the determination of the albedo of the interstellar grains that we present, than in any previous purported determination.

Our conclusion that the albedo of the interstellar grains is very low means that very little scattered ultraviolet 
starlight is present on the sky.  Yet many reports of detection of such radiation have appeared in the literature.  In his
Annual Review of Astronomy and Astrophysics 
review of cosmic diffuse ultraviolet background radiation, Henry (1991) was extremely skeptical of such claimed detections.  He
drew attention to only a single observation where he felt that there was ``convincing proof" that scattered ultraviolet
starlight has been detected, namely the observation reported by Hoberg (1990).  There are 
two elements to that proof, both of which are vital to the proof's validity.
The first is the presence of stellar absorption lines in Holberg's observed spectrum: so, the light {\em is} starlight.
 In particular, Lyman $\gamma$
is clearly seen, and Lyman $\beta$ is also present, although the wavelength seems slightly off.  The second element of proof is 
establishment of the {\em diffuse} character of the radiation.  The concern here is that the signal might merely be due to a faint star
in the spectrometer field of view.  Murthy et al. (1999), in analysing such {\em Voyager} spectra, tested carefully for the diffuse character of signal
by making use of spacecraft pointing jitter.  In the case of Holberg (1990), the observation involved a three degree scan
across the sky, through which
the signal persisted.  Holberg {\em was} observing the scattered light of some O-type star; the particular star has not been identified.  
(So, by the way, we can conclude that the 
albedo of the interstellar grains in the far ultraviolet, while low, is not {\em zero}).

In contrast, the other reviewer of diffuse ultraviolet background radiation 
in the same volume of the Annual Review of Astronomy and Astrophysics, Bowyer (1991), concluded that ``in the past ten years
remarkable progress has been made in our understanding of the diffuse ultraviolet background...we now know that this flux is primarily
galactic in origin."  The reviewer's error was his uncritical acceptance all published claims of detection of scattered starlight
simply on the basis of the authors' unsubstantiated assertions of such detection.

Despite this, Bowyer (1991) did  conclude that the albedo of the interstellar grains is low.  He arrived at the correct conclusion on
 the basis of one of the Berkeley (UCB) UVX observations (Hurwitz, Bowyer, and Martin 1991).  That observation and the corresponding Johns Hopkins
 University (JHU)
 UVX observations, are described by Murthy, Henry, and Holberg (1991).  There is a discrepancy between the JHU and UCB fluxes in the case
 of the observation that is critical to Bowyer's conclusion regarding the albedo,
  and I am strongly inclined to accept the UCB result over the JHU result, for
 reasons that are explained in Murthy, Henry, and Holberg (1991).  (Were one to accept, instead, the JHU result, Bowyer's conclusion would only
 be strengthened.)  Henry and Murthy (1994) criticized the basis of Hurwitz et al.'s conclusion, pointing out (quite correctly)
  that their analysis
 ignored the strong anisotropy in the radiation field that is illuminating the dust in question, but Bowyer's fundamental point is unexceptionable: 
 ``the resultant diffuse background at low Galactic latitudes was surprisingly faint and was only consistent with a low value for the
 albedo, independent of~$g$."
 
 While we have these important UVX observations before us, let me point out that in their paper, Murthy, Henry, and Holberg (1991)
 report {\em only an upper limit} of 100--200 \phunit at $\sim$ 1050 \AA\ from \voyager observation of
  this same, critical, UVX target.  Yet Hurwitz, Bowyer, and Martin (1991) reported
 $\sim$ 1000 \phunit in this direction, at wavelengths longward of Lyman $\alpha$.  This is the best {\em specific} example
  of measurement of the ``ledge" in the diffuse background, that 
 has been claimed by Henry (1991, 1999) to be {\em generally} present. In Figure~8, I present one example of a \voyager spectrum showing no
 diffuse radiation shortward of Lyman $\alpha$.  The reduction methodology for the \voyager data that was used
  by Holberg (1986) and by Murthy et al. (1999) has been 
 criticized by Edelstein et al. (2000), but Murthy et al. (2001) have demonstrated that those criticisms are invalid.  
 
 The existence
 of a significant diffuse ultraviolet flux at wavelengths longward of Lyman~$\alpha$ is {\em not} controversial (Henry 1991); such radiation has been
 detected most recently by Brown et al. (2000) using STIS on {\em HST}.  Those authors report a background of 501 $\pm$ 103 \phunit at a location
 with $E_{B-V}=0.01$.  The present determination of the albedo of the interstellar grains guarantees that this radiation is extragalactic. 
 Henry (1991, 1999) suggests that such radiation is recombination radiation from the intergalactic medium.

Finally, I discuss a paper in which Henry (1981) claimed that the albedo of the interstellar grains at 1565 \AA\ is {\em high} ($a>0.5$).  That
paper is very similar to the present paper: in that paper, the comparison was of these same \apollo observations with the integration
of the direct starlight that was observed with the TD-1 experiment (Gondhalekhar, Phillips, and Wilson 1980).
  The direct starlight radiation apparently fell short by about 25\%, and
Henry felt that the excess must be diffuse scattered ultraviolet light confined to low galactic latitudes.  Henry's error
was elementary:   he failed to include a well-understood
 correction for blended stars.  I now have integrated the entire set of TD-1 observations as they were
corrected for blended stars by Landsman (1984), finding a total flux of $1.252\times 10^{-6}$ \egunit.  In contrast, 
integrating the TD-1 stars that
appear, for example, in the CD that one can obtain from NASA's Goddard Space Flight Center yields only $1.041\times 10^{-6}$ \egunit.  
My computed increase
in brightness is in excellent accord with the similar correction that was presented by Gondhalekar, Phillips, \& Wilson (1980) 
themselves (their Table 5),
and brings the TD-1 observation into excellent accord with the present integrations, as we have already seen in Figure~4.
(I was annoyed, at the time, that Bowyer (1991) omitted reference to Henry (1981) in his review of determinations of the albedo; it 
now emerges that Bowyer's judgment on this point was excellent.)
\section{Conclusion}

In contrast to my first exploration (Henry 1977) of the use of star catalog integrations 
as a means of predicting the interstellar radiation field,
 I have not presented (or obtained) maps of the predicted distribution of the radiation on the sky.  Any
such predictions require knowledge of the shape of the scattering function of the interstellar grains, and are incomparably more
uncertain than is the evaluation of the integrated spectrum of the radiation field as a whole, on which I have here exclusively focused.

I conclude that the albedo of the interstellar grains in the far ultraviolet 
is very low, perhaps $a=0.1$.  I emphasize that the present determination is much better founded than any previous
determination by other methods.  The only significant uncertainty is the absolute calibration of the {\em Apollo 17}
ultraviolet spectrometer, but I have great confidence in the accuracy of that calibration.  In the present investigation I can
come to no independent conclusion concerning the value in the 
far ultraviolet of the Henyey-Greenstein scattering parameter $g$, but I am 
happy to accept the conclusion of Witt et al. (1992) that the value of $g$ is 
large, corresponding to very strong forward scattering.  Burgh (2001) reaches the same conclusion.  Finally,
I am able to conclude that Henry (1991, 1999) is correct in asserting that virtually all of 
the ultraviolet background radiation that is seen at high (and even at moderate) galactic latitudes is 
extragalactic in origin.  

\acknowledgments

This work was supported by NASA's Maryland Space Grant Consortium.  I thank Dr. E. Julius Dasch, who has done a
magnificent job in managing the National Space Grant College and Fellowship Program for the past 12 years.  Chet Opal
was a fine young man and an excellent scientist; I am very sorry that he died prematurely.

\appendix

\section{Interstellar Extinction}

In my work for this paper, I have used the interstellar extinction of Cardelli, Clayton, \& Mathis (1989, CCM).
Here,  I extract the various 
equations of CCM and assemble them to conveniently provide the average
 $A_\lambda/A_V$ as a function of two parameters: the inverse wavelength $x \equiv 1/\lambda ~\mu m^{-1}$, plus the value of 
the parameter $R_V = A_V/E_{B-V}$ that characterizes the shape of the extinction curve.

Infrared: \hspace{.5in}  $0.3~ \mu m^{-1} \le x \le 1.1~ \mu m^{-1}$ \hspace{.5in}   $(33,333~ \AA \ge x \ge 9091~ \AA)$

\begin{eqnarray*}  
A_{\lambda}/A_V=0.574 (x)^{1.61}-[ ~0.527 (x)^{1.61} ~]/R_{V} \nonumber\\
\end{eqnarray*}

Optical/NIR: \hspace{.5in}  $1.1~ \mu m^{-1} \le x \le 3.3~ \mu m^{-1}$  \hspace{.5in}  $(9091~ \AA \ge x \ge 3030~ \AA)$

\begin{eqnarray*}  
A_{\lambda}/A_V=1+0.17699 ({x}-1.82)-0.50447(x -1.82)^2-0.02427(x -1.82)^3 \nonumber\\
+0.72085(x -1.82)^4+0.01979(x -1.82)^5-0.77530(x - 1.82)^6+0.32999(x -1.82)^7 \nonumber\\
+[ ~1.41338(x -1.82)+2.28305(x -1.82)^2+1.07233(x -1.82)^3\nonumber\\
-5.38434(x -1.82)^4-0.62251(x -1.82)^5+5.30260(x -1.82)^6 \nonumber\\
-2.09002(x -1.82)^7 ~]/R_{V}\\
\end{eqnarray*}

Near Ultraviolet: \hspace{.5in}  $3.3~ \mu m^{-1} \le x \le 5.9~ \mu m^{-1}$ \hspace{.5in}   $(3030~ \AA \ge x \ge 1695~ \AA)$

\begin{eqnarray*}  
A_{\lambda}/A_V=1.752-0.316 ~x-0.104/[(x - 4.67)^2+0.341] \nonumber\\
+[ ~1.825 (x)-3.090 +1.206/[(x -4.62)^2+0.263]~ ]/R_{V}\nonumber\\
\end{eqnarray*}

Ultraviolet: \hspace{.5in}  $5.9~ \mu m^{-1} \le x \le 8~ \mu m^{-1}$ \hspace{.5in}  $(1695~ \AA \ge x \ge 1250~ \AA)$

\begin{eqnarray*}  
A_{\lambda}/A_V=1.752-0.04473 ({x}-5.9)^2-0.009779(x -5.9)^3 \nonumber\\
-0.316(x)-0.104/[(x - 4.67)^2+0.341] \nonumber\\
+[ ~0.2130(x -5.9)^2+0.1207(x -5.9)^3+1.825(x)\nonumber\\
-3.090 +1.206/[(x -4.62)^2+0.263 ]~]/R_{V}\nonumber\\
\end{eqnarray*}

Far Ultraviolet: \hspace{.5in}  $8~ \mu m^{-1} \le x \le 10~ \mu m^{-1}$ \hspace{.5in}   $(1250~ \AA \ge x \ge 1000~ \AA)$

\begin{eqnarray*}  
A_{\lambda}/A_V=-1.073-0.628 ({x}-8)+0.137(x -8)^2-0.070(x -8)^3 \nonumber\\
+[ ~13.670+4.257 ({x}-8)-0.420(x -8)^2+0.374(x -8)^3 ~]/R_{V} \nonumber\\
\end{eqnarray*}

A plot of these relations, for the two cases $R_V=3.1$ and $R_V=5$, is given in Figure 9.

\section{Observations of the Ultraviolet Radiation Field}

There are very few observations of the interstellar ultraviolet radiation field, because very fast
optics are required if the observation is to be sensitive to the diffuse component.
Apart from the {\em Apollo 17} observations that are described in the text, I know of only
two observations, both rocket measurements that were conducted from the United States Naval
Research Laboratory.  While I reject both of these observations on the grounds of
the uncertainty of their absolute calibration, I do wish to discuss them here, briefly, ``for 
the record."

Consider, first, the rocket observation by Henry, Swandic, Shulman, \& Fritz (1977) of the strength of 
the interstellar radiation field at 1530 \AA.
Those authors calibrated their detector before and after the rocket flight, but  
they found that observations of four stars during the flight
did not agree with those calibrations.  They chose to accept the inflight calibration, which resulted in a reduction of a factor 
1.67 $\pm$ 0.36 in their claimed intensity of the radiation field.  
However, I find (Table 1) that {\em if} those authors had assumed an albedo of
unity for the interstellar grains and strong forward scattering (and there {\em does} exist 
considerable evidence that the grains are
strongly forward-scattering in the far ultraviolet, for example Witt et al. 1992), then 
they would have attained an average agreement of ``observed/unreddened" of
1.03 (this being the average of the four values in the final column of Table 1), which is to 
say essentially perfect agreement between in-flight
and laboratory calibration.  

If we were to accept the Henry et al. rocket result as it was originally presented by the authors, we 
would conclude that comparison of it with
our integrations demonstrates a $2\sigma$ detection of 
forward-scattered starlight in the radiation field at 1530 \AA.  
On the other hand, if we
instead assume albedo unity, conjoined with strong forward 
scattering (which combination, we saw, 
produces the excellent agreement between their inflight and laboratory calibrations), comparison 
 with the present models suggests a value of the albedo of about $a=0.85$, and then re-calculation of the inflight calibration
(assuming $a=0.85$ rather than $a=1.0$), gives a $\sim 10\%$ disagreement
with the laboratory calibration, which is entirely acceptable.  

Finally, there is the rocket result of Opal and Weller (1984) at 975~\AA.  The Opal and Weller result, if it were to be 
accepted at face value, would indicate a value of the albedo of the interstellar grains, at 975~\AA, of about $a=0.6$.

While both these rocket results would imply a higher albedo for the interstellar grains than I have
found in this paper, I have no hesitation 
in accepting, instead, the {\em Apollo 17} observations, simply because of the much more meticulous
 instrumental calibration
that was made in the case of {\em Apollo 17}.

\clearpage


\begin{figure}
\plotone{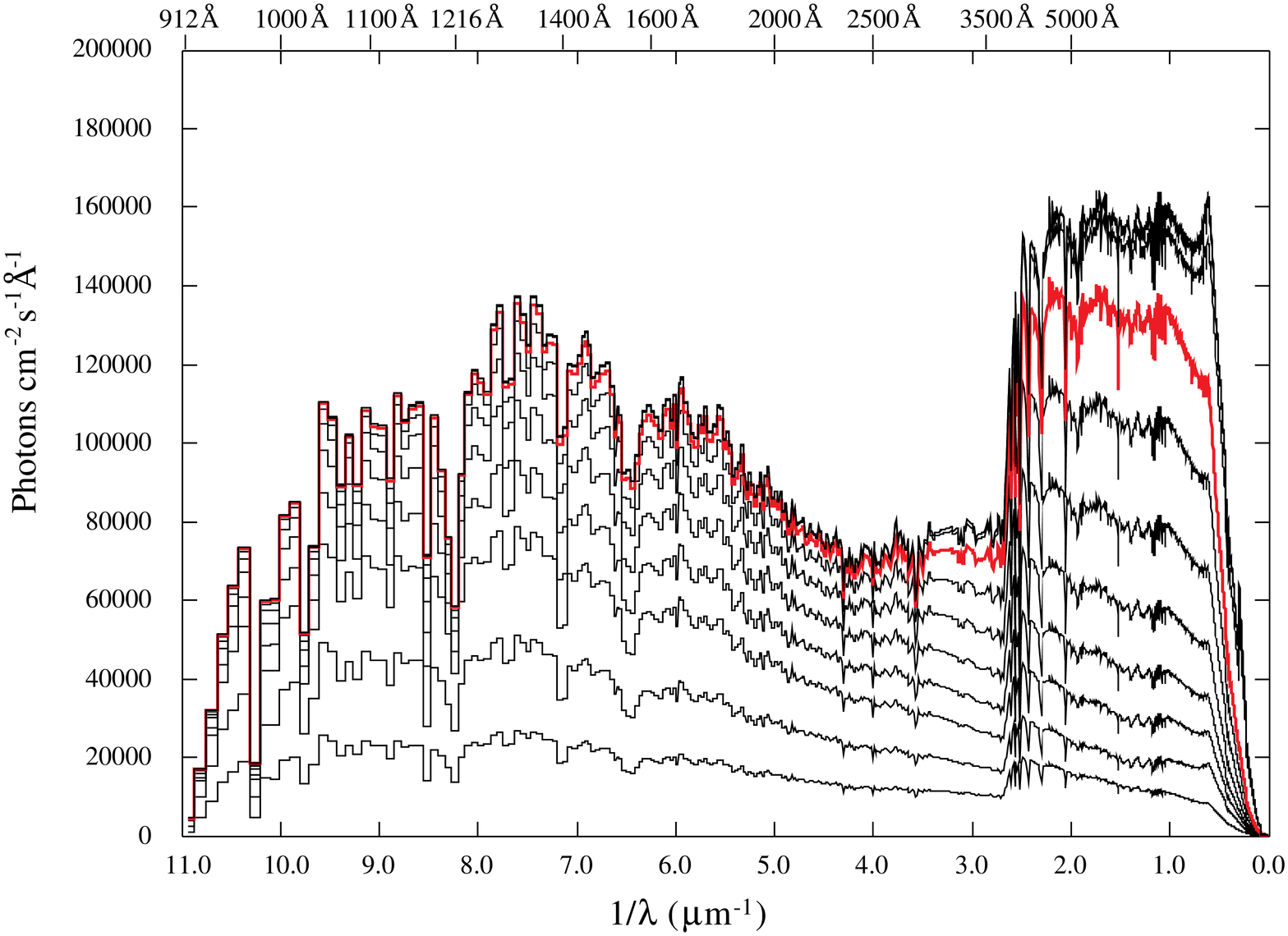}
\caption{The highest curve is the spectral distribution of the integrated radiation of all the stars in the
Hipparcos Input Catalog.  The intensity (in this first figure), is computed on the assumption 
that interstellar grains have an  albedo {\em a} of zero: the grains {\em absorb} 100\% of the extincted radiation.
  Also shown (lower curves) are the separate contributions to the total spectrum that are
provided by stars that are brighter than or equal to $V= 1.0, 2.0, 3.0, 4.0,
 5.0, 6.0, 7.0, 8.0, 9.0,$ and $10.0$ magnitudes.  The summation for the case ``brighter than or equal to
 $V=8.0$ magnitude" is drawn in red.
At the longer wavelengths (such as, for example, 3000 \AA\ $(3.33 ~\mu m^{-1})$  and 5000 \AA\  $(2.0 ~\mu m^{-1}))$, 
it is clear that the 
 catalog does not go deep enough to capture all contributors to the total radiation field.
 Note that the {\em apparent} convergence of the integration at the longest wavelengths (at a magnitude {\em fainter}
     than 8.0) is merely an artifact caused by incompleteness in the catalog.  
 In contrast, as first pointed
  out by Henry (1977), at far ultraviolet wavelengths, integration even to $V=8.0$ magnitude is 
   quite sufficient to gather all significant contributions to the radiation field---assuming that $a=0$.
   This figure demonstrates that this same conclusion holds for all wavelengths shortward of about 2800 \AA\ $(3.6 ~\mu m^{-1})$.
\label{fig1}}
\end{figure}
\clearpage %

\begin{figure}
\plotone{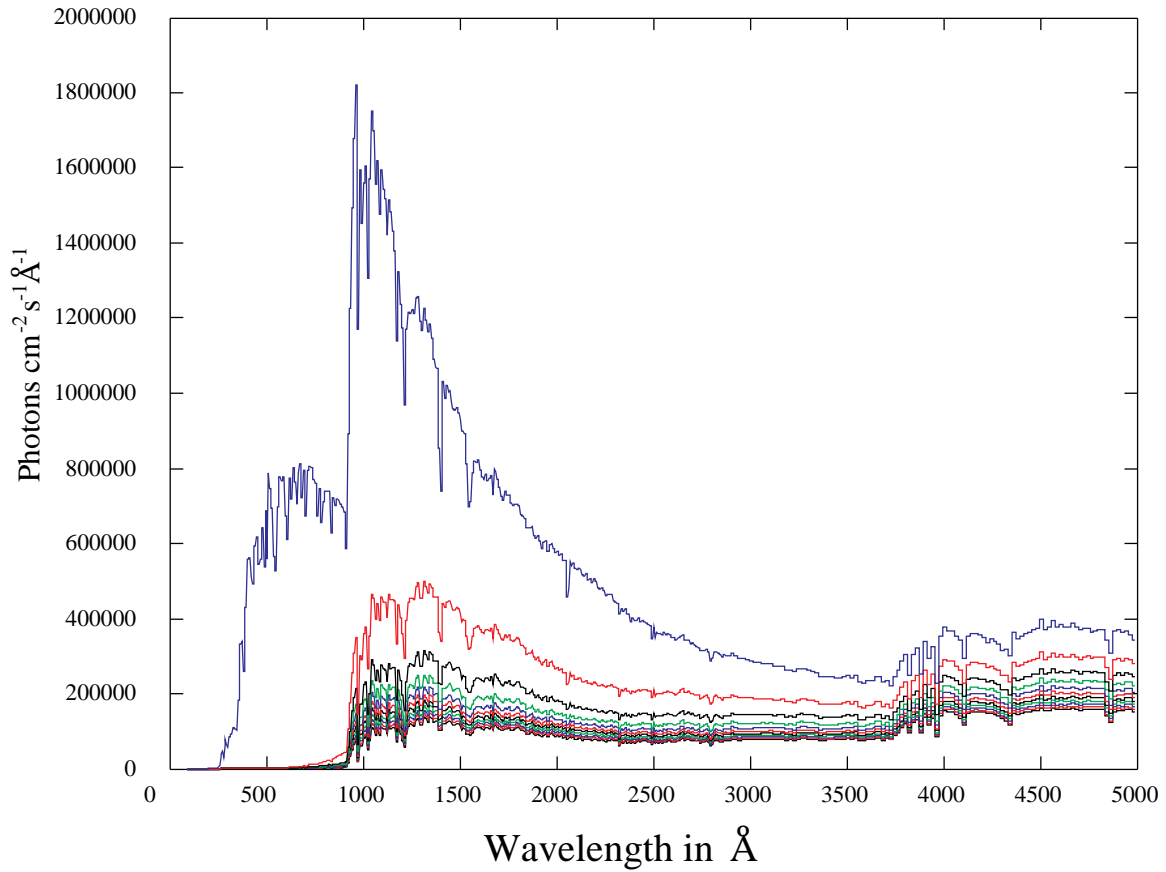}
\caption{The expected interstellar radiation field for the (realistic)
case that the extinction shape parameter $R_V = 3.1$, for values of the assumed interstellar grain albedo $a=0$ (lowest curve) 
to $a=1.0$ (fully reflective grains, uppermost curve).  Intermediate curves are in albedo steps 
of 0.1.  The lowest curve in {\em this} figure is the same as the {\em highest} curve of Fig. 1.  We
 see the potentially enormous effect on the interstellar radiation field of reflective interstellar grains.     
\label{fig2}}
\end{figure}
\clearpage

\begin{figure}
\plotone{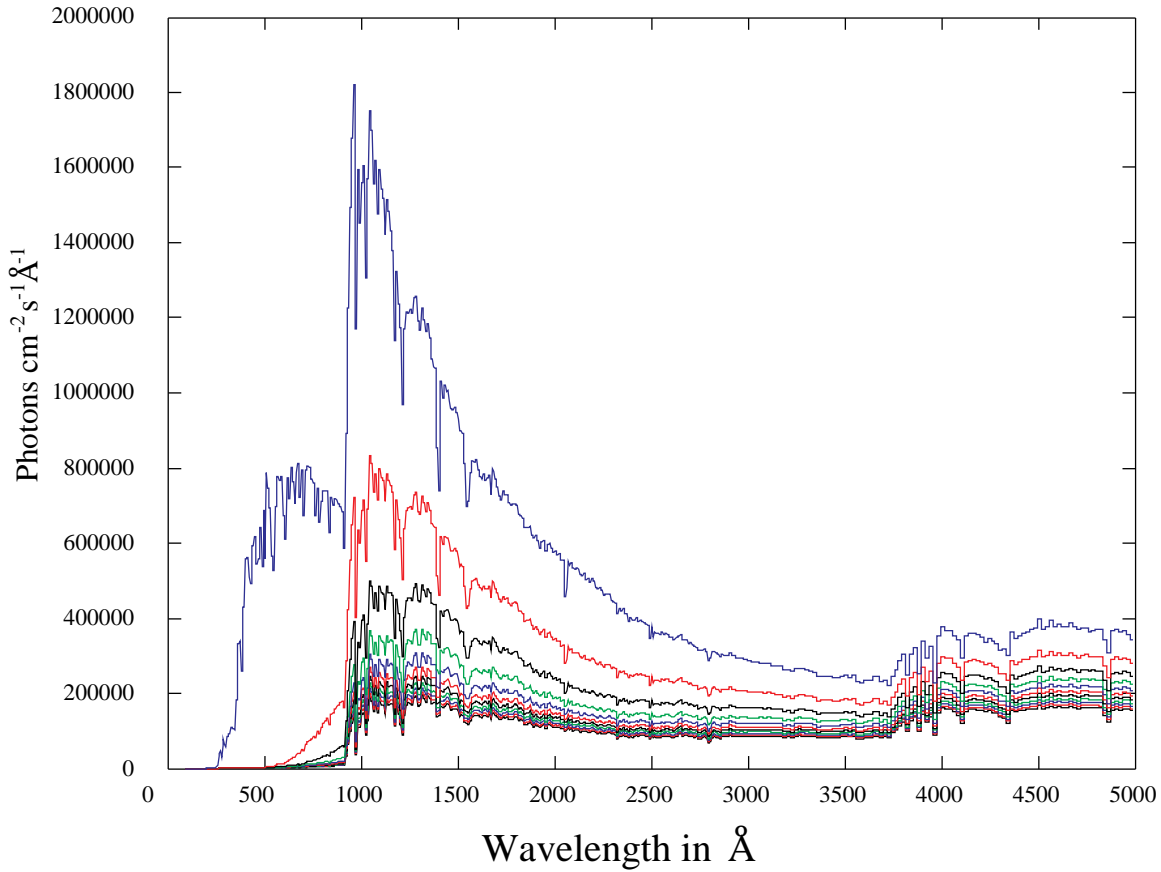}
\caption{This figure is the same as Figure~2, but shows the integration for the case $R_V = 5$, which is the value 
of $R_V$ that is found for dense interstellar
clouds.  The topmost curve is, of course, the same as in Figure~2, because when the albedo is unity, the 
grains remove {\em nothing} from the interstellar radiation
field, regardless of the strength of the interstellar extinction.  But the lowest curve in the present figure
is much {\em higher} than is the lowest curve in Fig.~3; when we compare with observations, we will find that this fact
allows us to arrive at the (expected) conclusion that the vast bulk of interstellar extinction arises from grains
that can be characterized by $R_V$ = 3.1, not by $R_V$ = 5. 
\label{fig3}}
\end{figure}
\clearpage

\begin{figure}
\plotone{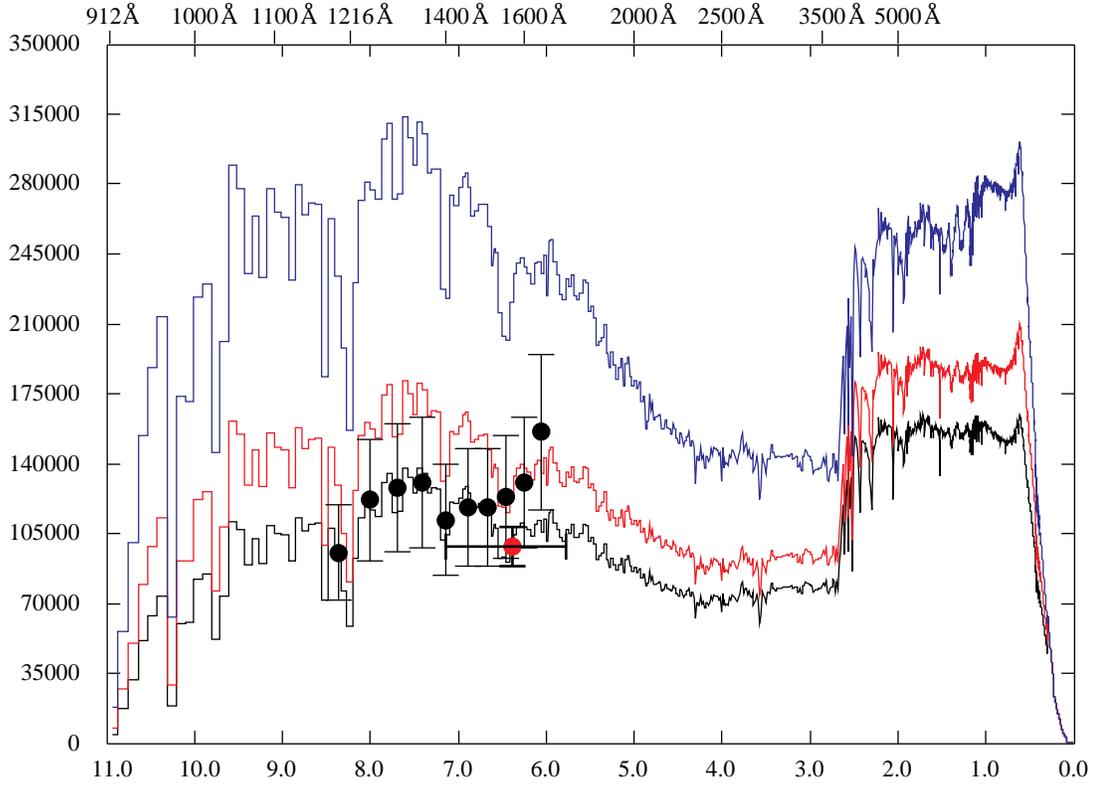}
\caption{An expanded view of part of Fig. 2 including only those curves corresponding to assumed values of the interstellar grain
 albedo of $a=0.0$, 0.4, and 0.8.  The \apollo  measurement of the spectrum of the interstellar radiation
  field of Henry et al. (1980) are the black circles.  Comparing these observations with the present models,
    I conclude that the albedo of the interstellar grains in the far ultraviolet is very low,
   perhaps $a=0.1$ in agreement with Murthy et al. (1991).  Longward of 1500 \AA\ ($\sim 6.7~\mu m^{-1}$)
   the albedo increases rapidly to $\sim 0.4$; at least,  I can think of no alternative explanation of these \apollo 
    data other than an increase in the grain albedo.  That this interpretation is correct is supported
   by consideration of the {\em other} observation that is included in this figure, the red circle that represents integration
   of all the stars that were observed at 1565 \AA\ using the TD-1 satellite (Gondhalekhar, Phillips, and Wilson 1980).  
   {\em That} observation, which includes {\em all} the TD-1 stars, but {\em no} diffuse radiation,
   matches my ``zero-albedo" integration, indicating that   the \apollo excess flux is surely scattered starlight. 
   \label{fig4}}
\end{figure}
\clearpage

\begin{figure}
\plotone{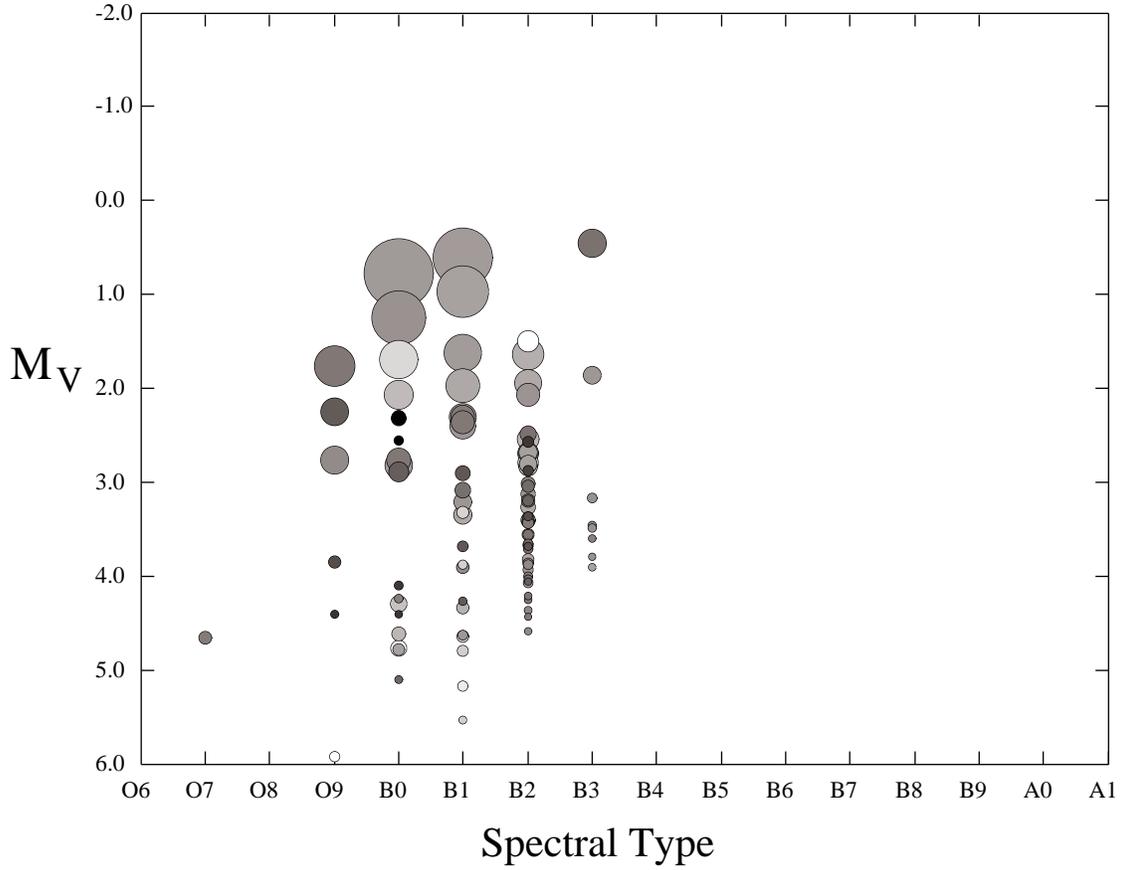}
\caption{The 100 stars that contribute the largest amount to the interstellar radiation field at 965 \AA.  The area of the 
 plotting symbol is proportional to the flux from the star at that wavelength.  The plotting symbol is 
 white if the star is unreddened; black if the star is the most heavily reddened of the 100 
 stars.  At 965 \AA\ the largest value of $A_V$ (among these 100 stars) is 0.229.  These 100 stars contribute
 87.5\% of the total flux at this wavelength (for the case that the interstellar grain albedo is assumed to be zero).
 \label{fig5}}
\end{figure}
\clearpage

\begin{figure}
\plotone{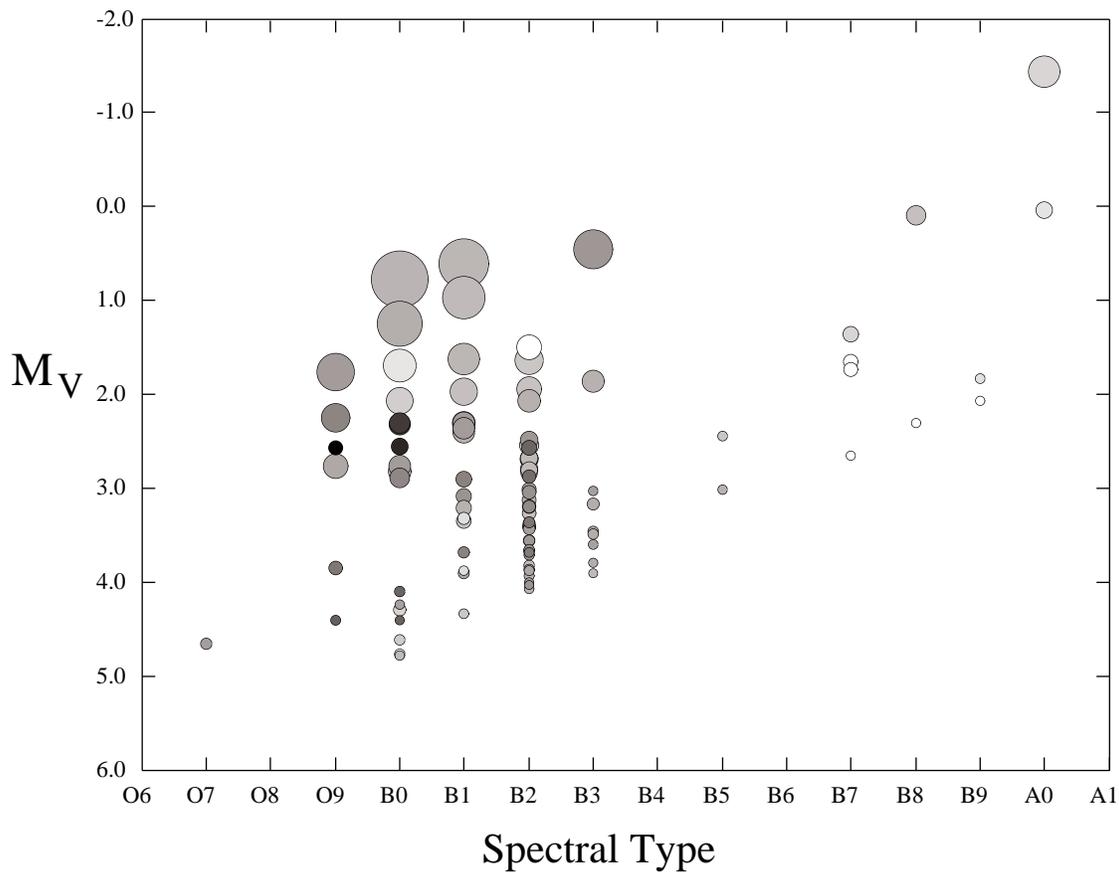}
\caption{The 100 stars that contribute the largest amount to the interstellar radiation field at 1535 \AA.  The area of the 
 plotting symbol is again proportional to the flux from the star at that wavelength.  The symbol is, again,
 white if the star is unreddened, and black if the star is the most heavily reddened of these 100 
 stars.  At 1535~\AA, the largest  $A_V$ among these 100 stars is 0.341. At this only slightly longer wavelength than that which 
 was analysed in Figure~5,
  the 100 brightest stars contribute
  only 62.1\% of the radiation field.  (The star at the top right is, of course, Sirius.)
  \label{fig6}}
\end{figure}
\clearpage

\begin{figure}
\plotone{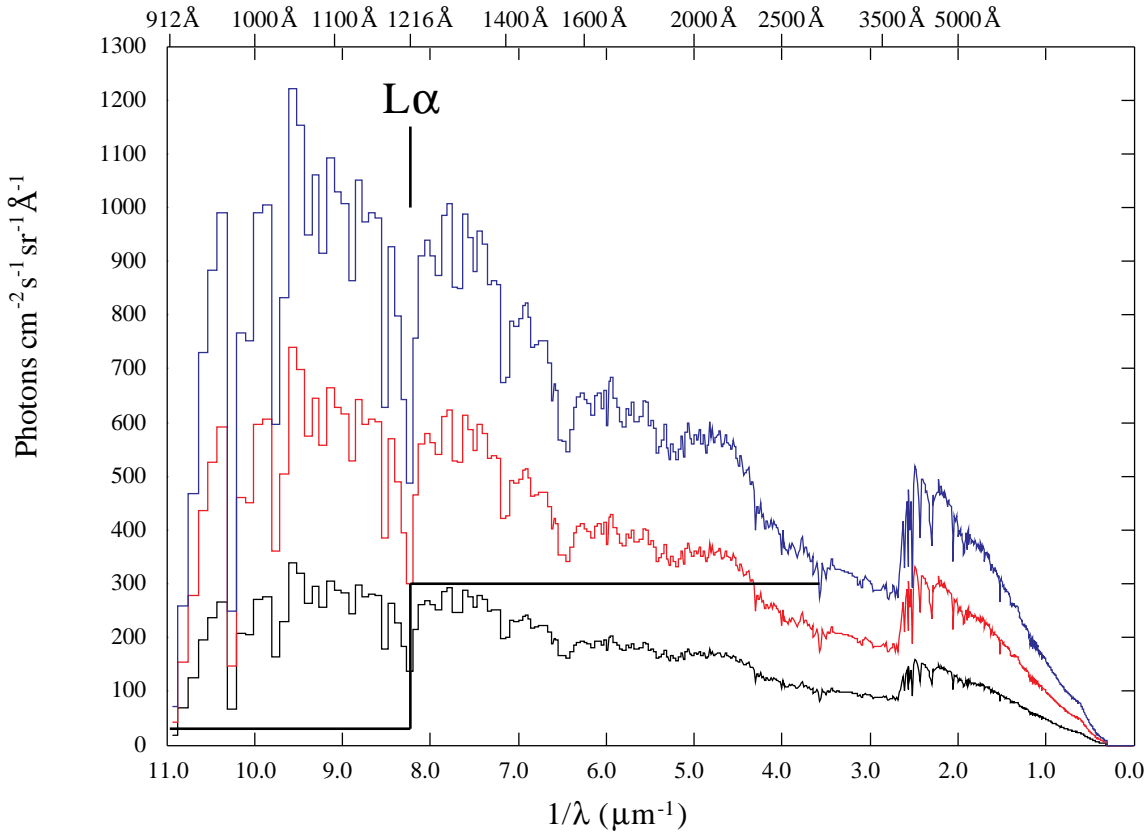}
\caption{The predicted background radiation at high galactic latitudes, that would originate in
the backscattered light from the interstellar radiation field that has been computed in the present
paper, if isotropic scattering is assumed, and dust corresponding to $A_V=0.1$ is present.
Also shown is a cartoon of the observations 
(straight black lines), showing the upper limit of Murthy et al. (1999) shortward
of Lyman $\alpha$, the abrupt rise at 1216 \AA\ that has been claimed by Henry (1999), and the 
broadly-accepted high-latitude intensity of perhaps 300 \phunit {\em longward} of L$\alpha$ that has been
reported by many observers:  the {\em non}-cartoon version of this information appears in Henry \& Murthy (1994).  
The lowest--intensity predicted spectrum corresponds to an albedo of 0.1, and agrees (crudely)
with the observations longward of Lyman~$\alpha$, but does {\em not} agree with the fact that Murthy
 et~al. find only an upper limit at  shorter wavelengths.  The higher-intensity curves
 represent backscattered radiation if the albedo is 0.2, or 0.3.  They disagree with the observations, but our model 
 assumes isotropic scattering.  It seems very likely that in the ultraviolet the grains 
 {\em strongly forward} scatter (Witt et al. 1992, Burgh 2001), in which case the predicted 
 intensities would be much lower than is shown in {\em any} of these spectra.  In
 any case, there are {\em no} circumstances in which the spectrum could 
 decline sharply below Lyman $\alpha$, relative to what is present {\em longward} of Lyman $\alpha$.  For example, even if the 
 grains were to become suddenly much more forward scattering below that wavelength (which would sharply change the
 scattered intensity), a similar phenomenon would necessarily 
 be seen in the case of the Coalsack nebula, and it is not (Murthy, Henry, and
 Holberg 1994).
 \label{fig7}}
\end{figure}
\clearpage

\begin{figure}
\plotone{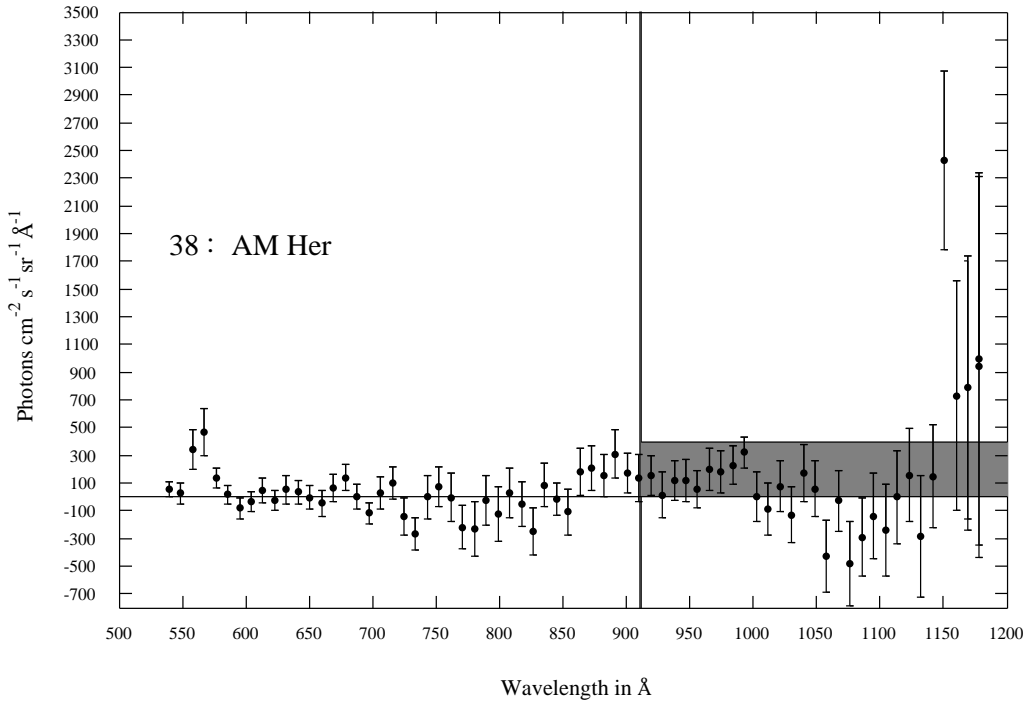}
\caption{A \voyager spectrum of a blank region of the sky at a high galactic latitude location near AM Her.  
The gray rectangle shows the level of diffuse ultraviolet background
radiation that typically occurs at wavelengths longward of Lyman $\alpha$.  We see from this \voyager spectrum that 
the upper limit on the diffuse background below Lyman $\alpha$ is well below 100 \phunit; the upper limit has been established by
Murthy et al. (1999) as 30 \phunit.  Henry (1999) has concluded from such data that the radiation that is observed longward
of Lyman $\alpha$ is redshifted recombination radiation from the ionized intergalactic medium.  This conclusion is greatly
strengthened by the present paper, which demonstrates that it is essentially impossible that the observed radiation at high
galactic latitudes could be the backscattered light of the general interstellar radiation field:  first, the albedo is too
low to provide such radiation, if $g$ is indeed high (as has been demonstrated by 
Witt et al. 1992, and by Burgh 2001), and second, if there {\em were} such radiation, the present models show that it
 must continue undiminished below Lyman~$\alpha$---but the observation that is presented in
this figure, for example, shows that it does not.
 \label{fig8}}
\end{figure}
\clearpage

\begin{figure}
\plotone{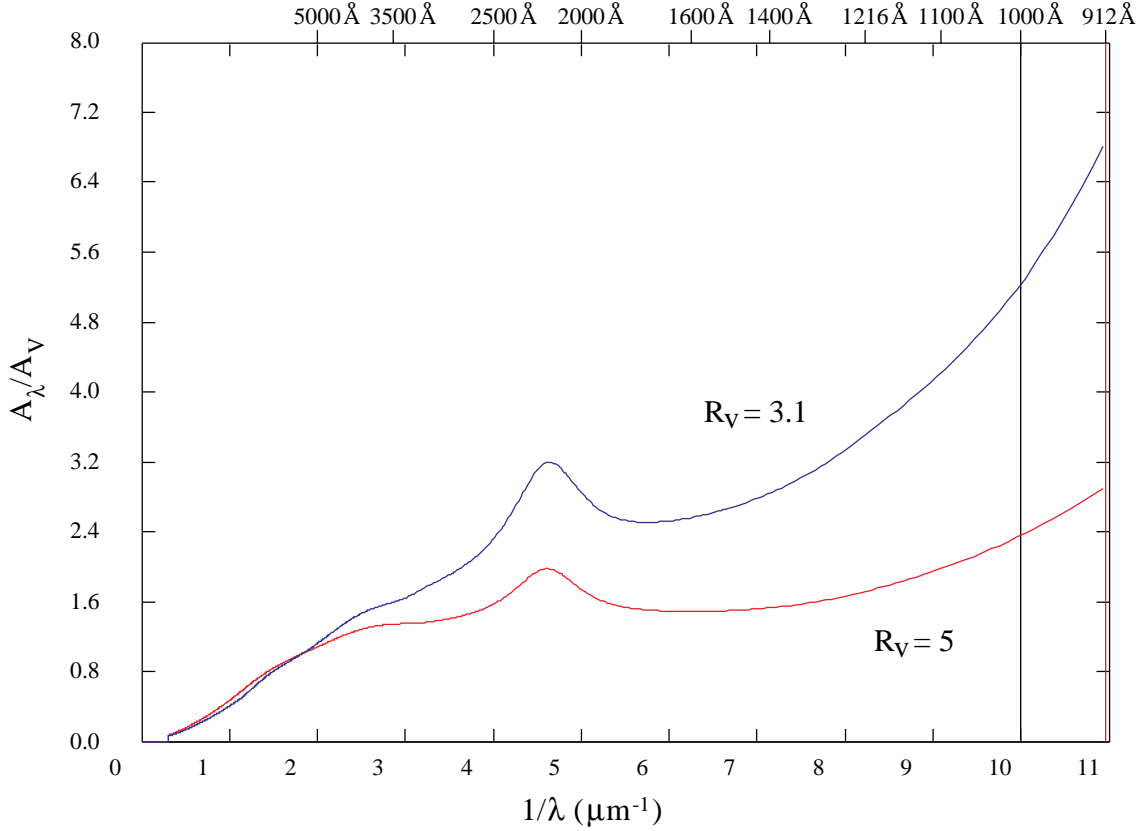}
\caption{This figure shows the two extinction curves (Appendix A) that were used in the present 
investigation.  They are due to Cardelli, Clayton, \& Mathis (1989, CCM).  Beyond 8   m$\mu$, CCM
warn that there is some uncertainty, but suggest that the curves have validity to 10 m$\mu$ (1000 \AA).
  I have placed vertical lines in the figure at 1000 \AA\ and at 912 \AA.  In the present paper, 
I have used CCM's formula right to the Lyman limit, as shown in the figure. (The extinction to the Lyman
limit has been investigated by Aannestad (1995), and does have the general character that appears in the present figure.) 
In the present work, I find that the albedo of the interstellar grains is $a\sim 0.1$ for inverse wavelengths
greater than $6.7~\mu m^{-1}$ ($\lambda <$ 1500 \AA) but is $a\sim 0.4$ for smaller values (longer wavelengths).  That
result suggests, from the present figure, that it is the {\em small} grains (those that are responsible for the large
{\em farthest} ultraviolet extinction) that have the very low albedo.  The same conclusion is reached by Burgh (2001)
on the basis of analysis of rocket observations of the reflection nebula NGC 2023.
\label{fig9}}
\end{figure}
\clearpage

\clearpage

\begin{deluxetable}{llcclccccc}
\tabletypesize{\scriptsize}
\tablecaption{Re-Calibration of the Rocket Observation of Henry et al. (1977). \label{tbl-1}}
\tablewidth{0pt}
\tablehead{
\colhead{Star} & \colhead{HR}   & \colhead{V}   &
\colhead{B-V} &
\colhead{Sp}  & \colhead{Observed} & \colhead{1977 adopted} &
\colhead{2001 model}     & \colhead{unreddend\tablenotemark{a}}  &
\colhead{${observed \over unreddened}$}
}

\startdata
$\gamma$ Cas &264 &2.47 &$-$0.15  & B0 IVe & 1875 &727  &805 &2053 &0.91 \\
$\alpha$ Leo &3982 &1.35 &$-$0.11 & B7 V   &370  &412  &400 &409 &0.91 \\
$\alpha$ Vir &5056 &0.98  &$-$0.23   & B1 III-IV  &4500  &3200  &2912 &4048 &1.12 \\
$\eta$ UMa &5191 &1.86 &$-$0.19 & B3 V  &1300 &750  &793 &1080&1.08\\
 \enddata

\tablenotetext{a}{This is the brightness for this star, if its V magnitude is corrected for reddening, but otherwise reddening is ignored.}

\tablecomments{There are so few observations of the interstellar ultraviolet radiation field that
each one must be considered carefully, even if contradictory results ultimately require rejection of some observations---such 
as this rocket observation.}

\end{deluxetable}

\end{document}